\documentclass[%
reprint,
superscriptaddress,
 amsmath,
 amssymb,
 aps, 
prb,
floatfix
]{revtex4-2}

\usepackage{graphicx}
\usepackage{dcolumn}
\usepackage{bm}
\usepackage{multirow}
\usepackage{threeparttable} 
\usepackage{float}
\usepackage{color}

\usepackage{hyperref}

\newcommand{\aog}{A$_\mathrm{1g}$}
\newcommand{\etg}{E$_\mathrm{2g}$}
\newcommand{\refsect}[1]{Sect. \ref{#1}}
\newcommand{\reftable}[1]{Table \ref{#1}}
\newcommand{\reffig}[1]{Fig. \ref{#1}}
\newcommand{\refeq}[1]{(\ref{#1})}

\usepackage{comment}

\begin{document}
\preprint{APS/123-QED}

\title{\textbf{Temperature-dependent Raman spectra of 2H-MoS$_2$ from Machine Learning-driven statistical sampling} 
}%

\author{Samuel Longo}
\affiliation{Nanomat group, Q-MAT, University of Liège,  and European Theoretical Spectroscopy Facility}%
\author{Alo\"{\i}s Castellano}
\affiliation{Nanomat group, Q-MAT, University of Liège,  and European Theoretical Spectroscopy Facility}%
\affiliation{Chemistry Department, Debye Institute for Nanomaterials Science and ETSF, Utrecht University, PO Box 80.000, 3508 TA Utrecht,
The Netherlands}
\author{Matthieu J. Verstraete}%
 \email{Contact author: Matthieu.Verstraete@uliege.be}
\affiliation{Nanomat group, Q-MAT, University of Liège,  and European Theoretical Spectroscopy Facility}%
\affiliation{ITP, Physics Department, Utrecht University 3508 TA Utrecht, The Netherlands}

\begin{abstract}
Molybdenum sulfides are in the spotlight of materials science thanks to their interesting properties for applications in optoelectronics, nanocomposites, lubricants, and catalysis.
The structural characterization of Molybdenum sulfides is a crucial step to understand and tune their properties. Vibrational techniques, such as infrared and Raman spectroscopy, can directly link to structural features, but the experimental literature suffers from large variability. Theoretical calculations are a powerful tool complementing and explaining empirical measurements. The reliability of first-principles calculation depends on the level of approximation made, taking into account disorder, doping, or temperature to yield a good description of the phonon statistics and related measurable quantities, such as the infrared and Raman peaks. 
In this study we calculate the Raman spectrum of crystalline 2H-MoS$_2$, including broadening and shifts due to thermal and anharmonic effects. Our results demonstrate excellent agreement with experimental measurements; notably, the calculated temperature trends in frequencies and linewidths align with empirical observations. These findings establish a robust computational framework, paving the way for similar studies on amorphous Molybdenum sulfides.
\end{abstract}

\maketitle

\section{Introduction}

Transition Metal Chalcogenides (TMCs) are promising materials for applications in optoelectronics \cite{Zheng2018, Mueller2018}, catalysis \cite{Jaramillo2007, Hinnemann2005, Khaidar2024, Fu2021}, as lubricants \cite{Vazirisereshk2019, Shankara2008}, in the polymer industry \cite{Papageorgiou2024, Yang1999, Ahmadi2020}, and even for elastic energy storage \cite{Peng2013}. 
Notably, Molybdenum Sulfides (MSs) exhibit promising catalytic activity in the Hydrogen evolution reaction \cite{Hinnemann2005, Luo2018, Zhu2019}, representing a non-toxic, cheap and less polluting alternative to current catalysts which are mostly based on Platinum and Lead \cite{Norskov2005, Cheng2016, McCrory2015}.
The tailoring of amorphous MSs catalysts, which show the highest activity  \cite{Benck2014, Merki2011}, is still strongly hindered by uncertainty on the actual electrolytic mechanism \cite{Parkinson2025, Tran2016}. This is also due to the absence of long-range order in these materials, which makes characterizing their structure a very hard challenge.

Raman spectroscopy is among the cheapest and most accessible characterization techniques, and is widely used to study TMCs \cite{Zhang2015, Rice2013, Mignuzzi2015}. As an example, in layered systems, the peak frequencies are extremely sensitive to the interlayer interaction, enabling quick identification of systems which are mono-, bi-layer, etc... \cite{Lee2010}. Experimentally many other effects can also lead to peak shifts, from strain and substrate epitaxy to dielectric environment and defects.

Our goal in the present work is to calculate the Raman spectrum of the simplest MS, 2H-MoS$_2$ in its bulk phase, including full thermal effects, to establish a baseline and dissect different contributions. Experimental measurements are performed at finite temperature, and spectral peaks will broaden and shift. Therefore, the most common 0 K Raman calculations \cite{Molina2011} are not enough, and a treatment of anharmonicity and temperature becomes necessary.

In the low-field regime, the most complete way to obtain the Raman spectrum is as the Fourier Transform of the polarizability autocorrelation function \cite{Gordon1965}. The latter can in principle be extracted from Molecular Dynamics (MD) simulations, but this requires calculating the dielectric tensor for each time step, making this route unviable due to its great computational cost.

An approximation to the Raman spectrum can also be calculated at the harmonic level with a few additional dielectric calculations. The quasi-harmonic approximation \cite{Allen2020} can be used to incorporate thermal expansion, but does not account for peak shifts due to explicit anharmonicity \cite{Yuan2015}. Furthermore, the shape of the Raman peaks, caused by the finite phonon lifetime and anharmonicity, is usually described as an empirical broadening of a Lorentzian bell. To capture both thermal expansion and anharmonicity, we rely on the Temperature Dependent Effective Potential \cite{Romero2015, Shulumba2017, Knoop2024, Hellman2013, Hellman-Abrikosov2013, Klarbring2020} (TDEP) method. Another element which is often missing in 0 K calculations is the quantum zero-point renormalization (ZPR); we investigate its effect by comparing two different sampling methods: classical MD (no ZPR) and a stochastic method using quantum occupations (with ZPR).
As TDEP itself requires a representative and accurate sampling of forces and positions (see \ref{subsec:IFC_extraction}), we leverage the Moment Tensor Potential \cite{Novikov2020} (MTP) Machine Learning Interatomic Potential (MLIP) to achieve accuracies close to DFT with a fraction of the cost. 

The rest of this paper is structured as follows: in Section \ref{sec:Methods} we present the computational methods that were used throughout the work. We start with the description of the MLIP and its training and testing procedure, followed by the sampling methods. Then we address the calculation of the vibrational properties within the TDEP framework, and, finally, the calculation of the Raman quantities.
In Section \ref{sec:results_and_discussion} we report and discuss our results. First we assess the performance of our MLIP, then we move to the determination of the unit cells at finite temperature. Next we investigate the phonon band dispersions and the spectral function properties, as compared to experiments. We conclude this Section presenting the Raman results. In the last Section, \ref{sec:conclusions}, we recapitulate and briefly present some future perspectives.

\section{\label{sec:Methods}Methods}
Our Raman workflow can be divided into distinct steps: (i) training an MLIP for MoS$_2$; (ii) sampling the temperature-dependent phase space; (iii) extracting vibrational properties, and (iv) calculating the Raman spectrum. In this section, a thorough explanation of each step is provided.

\subsection{\label{subsec:dataset}The MLIP}
We use the MTP framework, as it has proven to be an accurate MLIP for the calculation of vibrational and thermal properties of 2D and layered TMDs \cite{Marmolejo2022, Nair2025, Cui2023}. To generate atomic descriptors, MTP combines polynomials of the distance from the neighbors with n-th tensor powers of the relative position vectors, to include many-body features. Such combinations are contracted into scalars which are invariant under atomic permutations, reflections and rotations.

The ideal dataset used to train the MLIP should be representative of the structures the MLIP is meant to compute the properties of in production. In our case, we are interested in exploring the canonical phase space of bulk hexagonal MoS$_2$ between 100 and 700 K. For MTP (as for the majority of MLIPs), the only input to the model (except for the hyperparameters) is the chemical identity and the position of each atom in the system; therefore, we need realistic atomic positions in the above-mentioned temperature range, accounting for anharmonicity and phonon occupation. To obtain a collection of such structures (or configurations) we employed the Machine Learning Assisted Canonical Sampling algorithm \cite{Castellano2025, Castellano2022} (MLACS). 
The key idea is that starting from an initial structure (the 2H-MoS$_2$ crystal structure from Materials Project database \cite{Jain2013}, id: mp-2815) the ground state (GS) properties, such as energy, atomic forces and stress tensor, are computed via DFT (with Abinit \cite{Verstraete2025, Romero2020, Gonze2020, Vantroeye2016}) and used to perform a first training of the MLIP. With ML-driven Molecular Dynamics (MLMD) a short time evolution ($\sim$100 fs) (or trajectory) is simulated, to explore the phase space.
The GS properties and structures of selected configurations along the trajectory are stored in the dataset and the steps are repeated from the MLIP training. Thanks to the iterative dataset augmentation, the MLIP improves its description of the phase space regions already visited and therefore the MD simulation progressively gains in accuracy. 
To adequately cover the temperature-dependent extent of the phase space, at each MLACS iteration, a random temperature in the range is selected at every MLACS iteration for the MD simulation.
It is crucial, for the first MLACS iterations, to keep the simulations short, because the MLIP suffers from two quality-limiting factors: overfitting due to the small size of the dataset, and the early stage of the iterative procedure. As the MLIP gets better, it is safe to set longer simulation times.
To decide when to stop the loop, a convergence assessment strategy can be employed. We consider MLACS to be converged when the incorporation of new configurations into the dataset does not increase the fit error of the model trained on it, meaning that the new additions lie in a region already known by the MLIP. This is done by performing a k-fold cross-validation every N new configurations. For each dataset size, the error of the cross-validation is calculated as the average over the k folds, and used as our convergence criterion.

\subsection{\label{subsec:sampling}The sampling}
Before going through the sampling methods, it is useful to anticipate that the temperature-dependent vibrational properties, such as the interatomic force constants (IFCs), are extracted from the statistical correlations of the atomic forces of the system with the atomic positions, at a certain temperature. Hence, we are required to perform a temperature-dependent statistical sampling of the atomic forces and positions, meaning that the ensemble we collect must be representative of the system at the desired temperature. We compare two methods: stochastic sampling of an effective harmonic canonical ensemble and ML-MD simulations, running both samplings with a fully converged MLIP.

\subsubsection{\label{subsubsec:sTDEP}Stochastic sampling}
By making an initial guess of the IFCs, it is possible to determine the harmonic canonical ensemble (NVT) probability distribution function of the positions, which is easily shown to be a Gaussian distribution around the equilibrium geometry of each atom. Quantum phonon occupations can be enforced in the definition of the distribution function.
It is worth noting that only second-order IFCs (IFC2s - two-body interactions) are needed as we work in the harmonic canonical ensemble, and that they should be converged with respect to all numerical parameters (this is explained in the section dedicated to TDEP - see \refsect{subsec:IFC_extraction}).
The next step is to generate N random configurations out of this distribution and to compute the real (or reference) energy and forces (e.g. with DFT) and use them as an initial sampling to extract new IFCs.
Note that the calculation of the real properties with an accurate computational method ensures that anharmonic effects are included in the data (as average derivatives of the atomic forces with respect to the atomic displacements, i.e. IFCs).
The new effective IFC2s are extracted with TDEP; in this way they encompass the anharmonic effects mentioned above (see \ref{subsec:IFC_extraction}) and are renormalized to infinite order.
The previous steps are repeated until convergence.
We consider the sTDEP procedure to be converged if the full widths at half maximum (FWHM) of the Raman active modes' spectral function peaks  differ by less than 0.02 THz in successive iterations.
We start by generating 20 configurations, and we increase by 20 at each new iteration. 

\subsubsection{\label{subsubsec:MD_sampling}Molecular dynamics sampling}
While the stochastic sampling explores an effective ``harmonic canonical ensemble'', molecular dynamics simulations can also be used, to sample the full potential energy surface (PES) and the full (classical) canonical ensemble. Starting from the thermalized equilibrium (super)cell, the system is evolved under the action of the atomic forces and the constraints of a thermostat. Note that, as is often the case in the literature, not only the volume is fixed (NVT) but the full lattice vectors as well. After discarding the equilibration time steps, the rest of the trajectory can be used as sampling owing to the assumption of ergodicity, which allows one to replace thermal averages by time averages.
We perform 299 ps long simulations after 1 ps of equilibration (discarded), and sample forces and stress every 50 fs. 

Note that for both sampling methods, the equilibrium geometry at finite temperature must be defined, either by experiments (e.g. temperature-dependent XRD), numerically as NPT thermal averages, or via the quasi-harmonic approximation method. Our equilibrium cells are obtained as time averages of 40 independent NPT MD simulations with the same starting point, each lasting 95 ps (after 5 ps of equilibration, discarded) and sampled every 50 fs. To assess the convergence with respect to the number of instances, we compute differences of the $a$ and $c$ cell parameters obtained with 1, 2, 3 and 4 instances less. Then we do the same calculation for many shuffles of the instances and average the result. At any temperature, with 40 iterations we obtain average absolute differences below $5\cdot 10^{-6}$\AA. Finally, in both cases, the supercell is defined by expanding the hexagonal unit cell via a matrix of the shape:
$$
\begin{pmatrix}
m & m & 0\\
-m & m & 0\\
0 & 0 & n
\end{pmatrix}
$$
with $m$ and $n$ positive integers, yielding an orthorhombic cut of the lattice. Throughout this work, the notation $mmn$ will be used to refer to such supercells, though they are not ``diagonal'' extensions of the primitive unit cell.
\subsubsection*{}
It is important to note that both methods are formally justified, but differ substantially in the physics they comprise. MD-TDEP samples the exact classical canonical ensemble, accounting for all orders of anharmonic renormalization; this leads to exact static properties in the classical limit, but no zero-point motion is considered \cite{Hellman2013}. On the other hand, sTDEP relies on an effective harmonic sampling, needing some correction via free energy derivatives to properly account for the anharmonic renormalization; however, it allows for the enforcement of the Bose-Einstein statistics, which yields a quantum sampling including zero-point motion \cite{Bianco2017}.

\subsubsection{\label{subsubsec:convergence_supercell}Considerations on the supercell}
The choice of odd numbers of repetitions of the unit cell in the supercell along the in-plane and out-of-plane directions (such as 883) may appear unusual at first, but it is the result of a careful convergence study.

Given the limited computational resources, we choose to favor the in-plane convergence over the out-of-plane representation. However, in terms of diversity of the interactions, our choice is not worse than other possibilities with more repetitions along $c$. Indeed, the maximum cutoff allowed in TDEP calculations is determined by the radius of the biggest sphere inscribed in the supercell box. Given the high $c/a$ ratio of the unit cell, the maximum cutoff is usually limited by the in-plane (x-y) size of the supercell. In order to maximize the sampling efficiency, one should increase the number of repetitions along the smaller cell vectors, towards the cubic limit. The only advantage in choosing a largely non-cubic supercell, lies in the inclusion of more \emph{symmetry-equivalent} pairs and triplets, which improves statistical averaging. However, proper convergence should already be verified with respect to the size and variety of the sampling.

\subsection{\label{subsec:IFC_extraction}Extraction of the vibrational properties}
The temperature-dependent properties, such as the IFCs and phonon lifetimes, can be extracted using the TDEP method.
Introducing the convention of Latin letters for atoms and Greek letters for cartesian directions (unless stated otherwise), the potential energy surface is expanded as a Taylor series around the equilibrium geometry:

\begin{widetext}
\begin{eqnarray}\label{eq:PES}
    U(\mathbf{R-\mathbf{R}}_{eq})=U(\mathbf{u})
    =U(0)
    +\sum_{ij\alpha\beta}\left.\frac{1}{2}\frac{\partial^2 U}{\partial u_{i\alpha}\partial u_{j\beta}}\right|_{0}u_{i\alpha} u_{j\beta}
    +\sum_{ijk\alpha\beta\gamma}\left.\frac{1}{6}\frac{\partial^3 U}{\partial u_{i\alpha}\partial u_{j\beta}\partial u_{k\gamma}}\right|_{0}u_{i\alpha} u_{j\beta} u_{k\gamma}
    +...\nonumber\\
    =U(0)
    +\sum_{i\alpha}u_{i\alpha}\underbrace{\left[\sum_{j\beta}\frac{1}{2}\Phi_{i\alpha j\beta}^{(2)} \cdot u_{j\beta}\right]}_{\equiv F_{i\alpha}^{(2)}} 
    +\sum_{i\alpha}u_{i\alpha}\underbrace{\sum_{jk\beta\gamma}\left[\frac{1}{6}\Phi_{i\alpha j\beta k\gamma}^{(3)} \cdot u_{j\beta} u_{k\gamma}\right]}_{\equiv F_{i\alpha}^{(3)}} 
    +...
\end{eqnarray}
\end{widetext}
where $u_{i\alpha}$ is the displacement of atom $i$ along direction $\alpha$, $\Phi_{i\alpha j\beta}^{(2)}$ is the IFC2 between atom $i$ along $\alpha$ and atom $j$ along $\beta$, $\Phi_{i\alpha j\beta k\gamma}^{(3)}$ is the third-order IFC (IFC3). We also define the two-body force acting on atom $i$ along $\alpha$ due to the displacement of the other atoms as $F_{i\alpha}^{(2)}$ (and similarly for the 3-body forces).
Note that a first guess of the IFCs allows to initialize the model, and yields a tentative calculation of the atomic forces. The IFC2s are obtained by minimizing the average difference between the reference forces and those computed with the model. The higher-order IFCs are calculated sequentially on the residual of the previous orders. For example, the IFC3s are obtained by minimizing the error in the residual forces after removing the second-order part.

The first-order terms in equation \refeq{eq:PES} (not shown) are zero at the equilibrium geometry by construction. However, the finite-temperature NVT dynamics can produce differences in the average atomic positions. In this case, the NPT cell (alone) with the 0 K reduced coordinates will not yield the equilibrium geometry, and the perturbative approach does not hold, leading to non-zero first-order terms. To overcome this problem, we fit the first-order terms as well and displace (via gradient descent) the atoms towards new positions. By repeating this operation iteratively, the atomic reference positions converge to a minimum-energy geometry \cite{Hellman2013}.

The spatial disproportionation of the electric charge in polar materials leads to degeneracy lifting between longitudinal and transverse optical modes (LO-TO splitting). This must be taken into account when computing phonons close to $\Gamma$, by calculating correction terms to the dynamical matrix of the shape:
\begin{equation}
    \Delta\tilde D_{\alpha\beta}^{i,j}(\mathbf{q}\rightarrow 0 )\propto
    \frac{(\sum_\gamma q_\gamma Z_{i\alpha\gamma}^*)(\sum_\gamma q_\gamma Z_{j\beta\gamma}^*)}{\sum_{\gamma\delta}q_\gamma\varepsilon^\infty_{\gamma\delta}q_\delta}.
\end{equation}
Here $Z^*_{i\alpha\gamma} = \partial U^2/\partial u_{i\alpha}\partial\mathcal{E}_\gamma$ is the Born-effective charge associated to displacement of atom $i$ along $\alpha$, producing an electrical polarization along $\gamma$. $q_\gamma$ is the $\gamma$ component of the phonon wavevector $\mathbf{q}$ and $\varepsilon^\infty_{\gamma\delta}$ is the dielectric susceptibility along $\delta$ due to an electric field along $\gamma$.

Assuming random distribution of isotopes across the lattice, Tamura's model \cite{Tamura1983} is used to account for the effect of mass disorder on the spectral function. The correction to the imaginary part of the spectral function of phonon $\lambda$ is given by:
\begin{equation}
    \Gamma_\lambda^{\mathrm{iso}}(\omega)=
    \frac{4}{\pi}\omega_\lambda
    \sum_{\lambda'}\left[
    \omega_{\lambda'}
    \delta(\omega-\omega_{\lambda'})
    \sum_i g_i\left|\boldsymbol{\xi}_i^{\lambda\dagger} \cdot\boldsymbol{\xi}_i^{\lambda'}
    \right|^2\right]
\end{equation}
where the external sum runs over the phonon modes, $\boldsymbol{\xi}_i^\lambda$ is the atom-$i$ component of the eigenvector associated to the phonon $\lambda$, and the mass variance parameter $g_i$ for atom $i$ is given by:
\begin{equation}
    g_i=\sum_\kappa c_i^\kappa\left(
    \frac{M_i^\kappa - \bar M_i}{\bar M_i}
    \right)^2
\end{equation}
    while $c_i^\kappa$ and $M_i^\kappa$ are, respectively, the fractional concentration and mass of isotope $\kappa$ of atom $i$, and $\bar M_i=\sum_\kappa c_i^\kappa m_i^\kappa$.

At each order, the interatomic force constants are limited by a cutoff radius, with respect to which the IFCs must be converged.

For IFC extraction we first optimize the unit cell (up to a maximum atomic displacement of $0.0001$ \AA{}) and then converge the spectral function for the modes of interest with respect to the cutoff radii.

We converge the spectral functions with a fixed third-order cutoff radius (rc3) of 5 \AA{}, and varying the second-order cutoff radius (rc2). With the converged rc2 value of 12.35 \AA{}, we then converge the FWHMs and peak frequencies below 0.005 THz with $\mathrm{rc3}=8$ \AA{}.

For a given phonon mode $\lambda$, the spectral function $A_\lambda(\omega)$ measures the distribution of the states due to the broadening arising from scattering between phonons; it is proportional to the imaginary part of the phonon Green's function \cite{Cowley1963}:
\begin{eqnarray}
    A_\lambda(\omega) & = & -\frac{1}{\pi}\Im\left[G_\lambda(\omega)\right] \nonumber \\
    & = & \frac{1}{\pi}\frac{\Gamma_\lambda(\omega)}{\left[\omega^2 - \omega_\lambda^2 - \Delta_\lambda(\omega)\right]^2 + \Gamma_\lambda^2(\omega)}
\end{eqnarray}
where $\Delta_\lambda(\omega)$ and $\Gamma_\lambda(\omega)$ are, respectively, the real and imaginary parts of the self-energy. Note that according to this definition the self-energy has units of squared frequency, but TDEP follows the convention of normalizing it by a factor of $2\omega_\lambda$: $\Delta_\lambda(\omega)^{\mathrm{TDEP}} \propto \Delta_\lambda(\omega)/\omega_\lambda$ (and similarly for $\Gamma$).
The imaginary part of the self-energy is computed from the IFC3s as \cite{Romero2015}:
\begin{eqnarray}\label{eq:scattering_rate}
    &&\Gamma_\lambda^{(3)}(\omega) \propto \sum\limits_{\lambda'\lambda''}|\Phi_{\lambda\lambda'\lambda''}|^2
    \left\{(n_{\lambda'}+n_{\lambda''}+1)\delta(\omega-\omega_{\lambda'}-\omega_{\lambda''})
    \right.\nonumber\\
&&+\left.(n_{\lambda'}-n_{\lambda''})
\left[\delta(\omega-\omega_{\lambda'}+\omega_{\lambda''})-\delta(\omega+\omega_{\lambda'}-\omega_{\lambda''})\right]\right\}
\end{eqnarray}
which represents a sum over all three-phonon scattering events with $\Phi_{\lambda\lambda'\lambda''}$ being the IFC3s in phonon space, and $n_\lambda$ being the Bose-Einstein occupation. The real part is obtained via Kramers-Kronig transformations.  
We converge the error on the FWHMs below 0.005 THz for the 64x64x64 summing grid, with respect to the 56x56x56 case. The choice of the resolution of the frequency axis is also critical, especially when the spectral function is used to compute non-integrated quantities (such as the Raman spectrum). In such cases, the local shape of the peaks is determinant and it can be dramatically affected by low resolution. For our spectral functions, we converge to a resolution below 0.0034 THz.

\subsection{\label{subsec:Raman_theory}Calculation of the Raman spectrum}
The first-order Raman intensity is quantified by the first-order response of the dielectric function with respect to the atomic vibration at the equilibrium geometry, $\partial\chi/\partial Q$. 

For the Stokes scattering, the efficiency (proportional to the acquired spectrum) for a phonon $\lambda$ is given by \cite{Cardona1982}:
\begin{eqnarray}\label{eq:raman_scattering_eff}
    S_\lambda(\omega) & = &
    \frac{\hbar(\omega_e-\omega)^4}{32\omega_\lambda\pi^2c^4}
    \left|
    \mathbf{\hat e}_s\cdot\
    \frac{\partial\chi'}{\partial \mathbf{Q}_\lambda}\cdot
    \mathbf{\hat e}_i
    \right|^2 \cdot\nonumber
    \\&& \cdot
    \left[n(\omega_\lambda, T)+1\right]  
    \tilde A_\lambda(\omega)
\end{eqnarray}
where $\omega_e$ is the frequency of excitation, $\omega$ is the Raman shift, $\omega_\lambda$ is the harmonic frequency of the phonon, and $\mathbf{\hat e}_s$ and $\mathbf{\hat e}_i$ are the polarization vectors of the incoming and scattered light, respectively. With $n(\omega_\lambda, T)$ we denote the occupation of the phonon mode at temperature $T$ (excluding the zero-point-motion), while $\tilde A_\lambda(\omega)$ is the spectral function, normalized to $1$. Finally, $\partial\chi'/\partial Q_\lambda$ is the intensive Raman tensor:
\begin{eqnarray}\label{eq:raman_tensor_atomic_deriv}
    \frac{\partial\chi'}{\partial \mathbf{Q}_\lambda} & = &
    \sqrt{V_u}\sum_j^{\mathrm{ucell}}(M_j)^{-1/2}\frac{\partial\chi}{\partial\mathbf{u}_j}\cdot\boldsymbol{\xi}_j^\lambda\nonumber\\
    & = & \sqrt{V_u}\sum_j^{\mathrm{ucell}}(M_j)^{-1/2}\sum_\gamma^{\mathrm{cart}}\frac{\partial\chi}{u_{j\gamma}}\xi_{j\gamma}^\lambda
\end{eqnarray}

where $V_u$ is the volume of the unit cell, $M_i$ is the mass of atom $i$, ${\partial\chi}/{\partial\mathbf{u}_j}$ is a rank-3 tensor, whose element $(\alpha,\beta,\gamma)$ is ${\partial\chi_{\alpha\beta}}/{\partial u_{j\gamma}}$, and $u_{j\gamma}$ is the atomic displacement of atom $j$ along the cartesian direction $\gamma$. Finally, we denote by $\chi_{\alpha\beta}=\partial^2 U/(\partial \mathcal{E}_\alpha\partial \mathcal{E}_\beta)$ the change in the polarization along $\alpha$ to an electric field along $\beta$, namely the static dielectric tensor. Note that we use the stable-isotope average for the mass (as from \texttt{ASE.data.atomic\_masses}, in the Atomic Simulation Environment python package \cite{Larsen2017}).

It is important to specify that \refeq{eq:raman_scattering_eff} is valid for eigenvectors of the harmonic Hamiltonian defined with the position operator $Q_\lambda\propto1/\sqrt{2\omega_\lambda}(a^\dagger_{-\mathbf{q}_\lambda}+a_{\mathbf{q}_\lambda})$, which produces the $1/\omega_\lambda$ factor in the equation. Note that according to \refeq{eq:raman_scattering_eff} the spectrum given by the normalized spectral function is multiplied by a constant of the Raman shift (but its value depends on the phonon frequency), if we consider $\omega_e-\omega \simeq \omega_e$. 

The calculation of the Raman tensor is the most time-consuming step. One way to accurately perform it is with Density Functional Perturbation Theory \cite{Baroni2001, Gonze1997} (DFPT); in this case the third mixed derivatives must be computed, which requires considerable resources. In order to save computational time, we displace each atom in the unit cell along each cartesian direction by a small quantity (in both positive and negative directions, to reduce noise), and compute the dielectric tensor via DFPT for the displaced unit cells. By finite-differences we obtain the variation of the polarizability with respect to atomic displacements, which can be used to write the Raman tensor via the phonon eigenvectors, according to \refeq{eq:raman_tensor_atomic_deriv}.

Note that, especially for the back-scattering geometry, most common in experiments, the conservation of momentum implies that only very-long-wavelength phonons are excited, hence only modes very close to the center of the Brillouin Zone are probed. For this reason, we only consider phonons at $\Gamma$, as it is commonly done. 

\subsection{Note on polarization}
Equation \refeq{eq:raman_scattering_eff} is valid for a given pair of polarization vectors of incoming and scattered light. Most Raman measurements on MoS$_2$ are done on powder samples, and hence we need to make an isotropic average over the possible polarizations. This is done in many textbooks, but usually only for cubic symmetry. We find the derivations of Ref.~\cite{Derek2002} to be general and clear. For unpolarized exciting and scattered light: 

\begin{eqnarray}\label{eq:iso_average_1}
 \left\{ \left|
\mathbf{\hat e}_s\cdot\
\frac{\partial\chi'}{\partial \mathbf{Q}_\lambda}\cdot
\mathbf{\hat e}_i
\right|^2 \right\} _{avg}\equiv
I_{avg} 
\propto 45\alpha'^2 + 7 \beta'^2
\end{eqnarray}
with
\begin{eqnarray}
    \alpha' & = & \frac{1}{3}\tilde\alpha'_{xx} + \tilde\alpha'_{yy} + \tilde\alpha'_{zz}\nonumber\\
    \beta'^2 & = &\frac{1}{2}\left[
    6(\tilde\alpha'^2_{xy}+\tilde\alpha'^2_{xz}+ \tilde\alpha'^2_{yz}) + \right. \nonumber\\
       &+&  \left. (\tilde\alpha'^2_{xx}-\tilde\alpha'^2_{yy})^2 
       +(\tilde\alpha'^2_{xx}-\tilde\alpha'^2_{zz})^2  
       +(\tilde\alpha'^2_{yy}-\tilde\alpha'^2_{zz})^2 
                   \right]\nonumber \\
    \tilde\alpha'_{\alpha\beta} &\equiv&\frac{\partial\chi'_{\alpha\beta}}{\partial\mathbf{Q}}
\end{eqnarray}
(we omit the phonon index in the last equations for clarity).

In order to account for the LO-TO splitting for a powder, a spherical integration should also be performed over the possible incoming light propagation directions (on top of the polarization direction average). Since LO-TO splitting is particularly small for MoS$_2$ (the calculated Born effective charges are $\sim$$1e$ and $\sim$$0.5e$ for Mo and S respectively), 
we do not consider averaging over the incident light direction, but only over the Raman tensor components for powders.

\section{\label{sec:results_and_discussion}Results and discussion}

In this section we first evaluate the performance of the MLIP, then we present the results of the TDEP-DFPT approach for the calculation of the Raman spectrum with both MD-based and stochastic sampling. Our analysis focuses on the evolution in temperature of the Raman active modes \aog{} and \etg, with particular attention to phonon frequency shift and spectral broadening.

\subsection{\label{subses:MLIP_performance}The MLIP training}
We train the MTP model with the following hyperparameters (see Ref.~\cite{Novikov2020}): \{level: 20, cutoff radius: 8 \AA, radial basis size: 8, radial basis type: Chebyshev polynomials; weights: (1,1,1)\}. For the k-fold cross-validation, the average root mean squared errors (rmse) on energy, forces and stress are below 0.003 eV/atom, 0.022 eV/\AA{} and 0.0005 eV/\AA$^3$, respectively, with a dataset size of 339 structures with 96 atoms each. The final training (with the same hyperparameters) is performed on the entire dataset and achieves rmse and R$^2$ of 0.00176 eV/atom, 0.01796 eV/\AA{}, and 0.00035 eV/\AA$^3$ on the energy, forces and stress, respectively. As a further confirmation of our MLIP being able to align to DFT, we use the MLIP to perform a structure relaxation with LAMMPS \cite{Thompson2022}. The optimized cell parameters differ by less than 0.00012 \AA{} and 0.027 \AA{} for $a$ (and $b$) and $c$, respectively. 

\subsection{\label{subsec:ucell determination}Thermalized unit cells}

\begin{figure}[thb]
    \centering
    \includegraphics[width=\columnwidth]{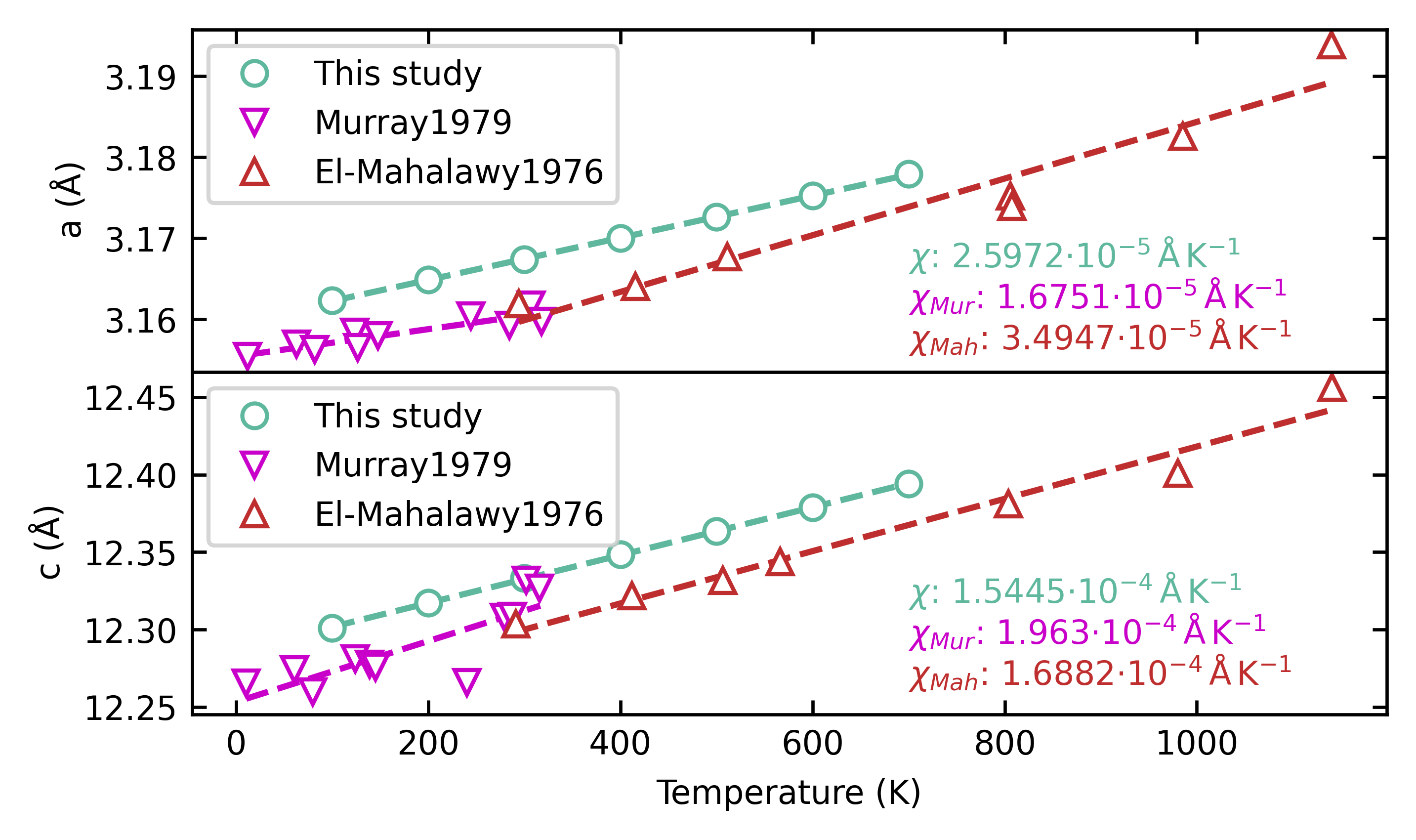}
    \caption{2H-MoS$_2$ lattice parameters vs temperature from our study and experimental work by El-Mahalawy et al. \cite{El-mahalawy1976} and Murray et al. \cite{Murray1979}. The data was fit to a linear function in T (dashed lines), and $\chi$ are the angular coefficients.}
    \label{fig:latt_pars_vs_T_exp_vs_theo}       
\end{figure}

\begin{figure}[thb]
    \centering
    \includegraphics[width=\columnwidth]{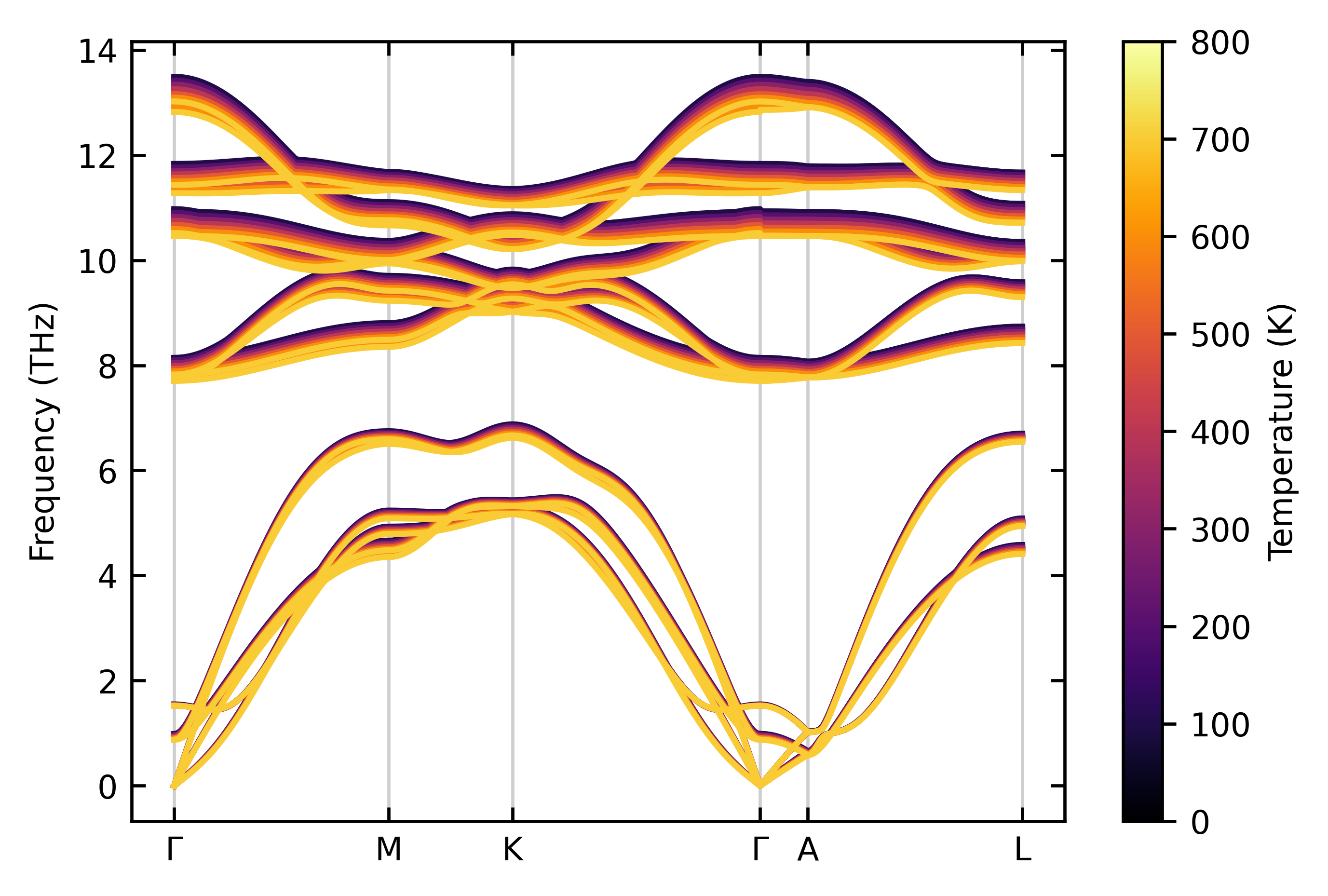}
    \caption{Phonon bands of MoS$_2$ from 100 to 700 K calculated with MD-TDEP.}
    \label{fig:phonons_vs_T}       
\end{figure}

We determine the lattice parameters of the thermalized unit cell between 100 and 700 K, according to the method described in \refsect{subsubsec:sTDEP}, using LAMMPS for the MD simulations, as in the rest of the study. In \reffig{fig:latt_pars_vs_T_exp_vs_theo} the variation in the lattice parameters with temperature is shown, compared to the experimental measurements from Ref.~\cite{El-mahalawy1976, Murray1979}. Our calculated parameters are in excellent agreement with the experimental ones, overestimating by only $\sim$$0.16\% $ and $\sim$$0.2\%$ respectively for $a$ and $c$, which is within the typical accuracy of GGA calculations with van der Waals corrections for this kind of layered systems \cite{Gusakova2017, Espejo2013}.
A good estimation of the lattice constants is crucial to ensure that the model for the PES, and therefore any derived quantity, is meaningful. 

\begin{figure}[thb]
    \centering
    \includegraphics[width=0.9\columnwidth]{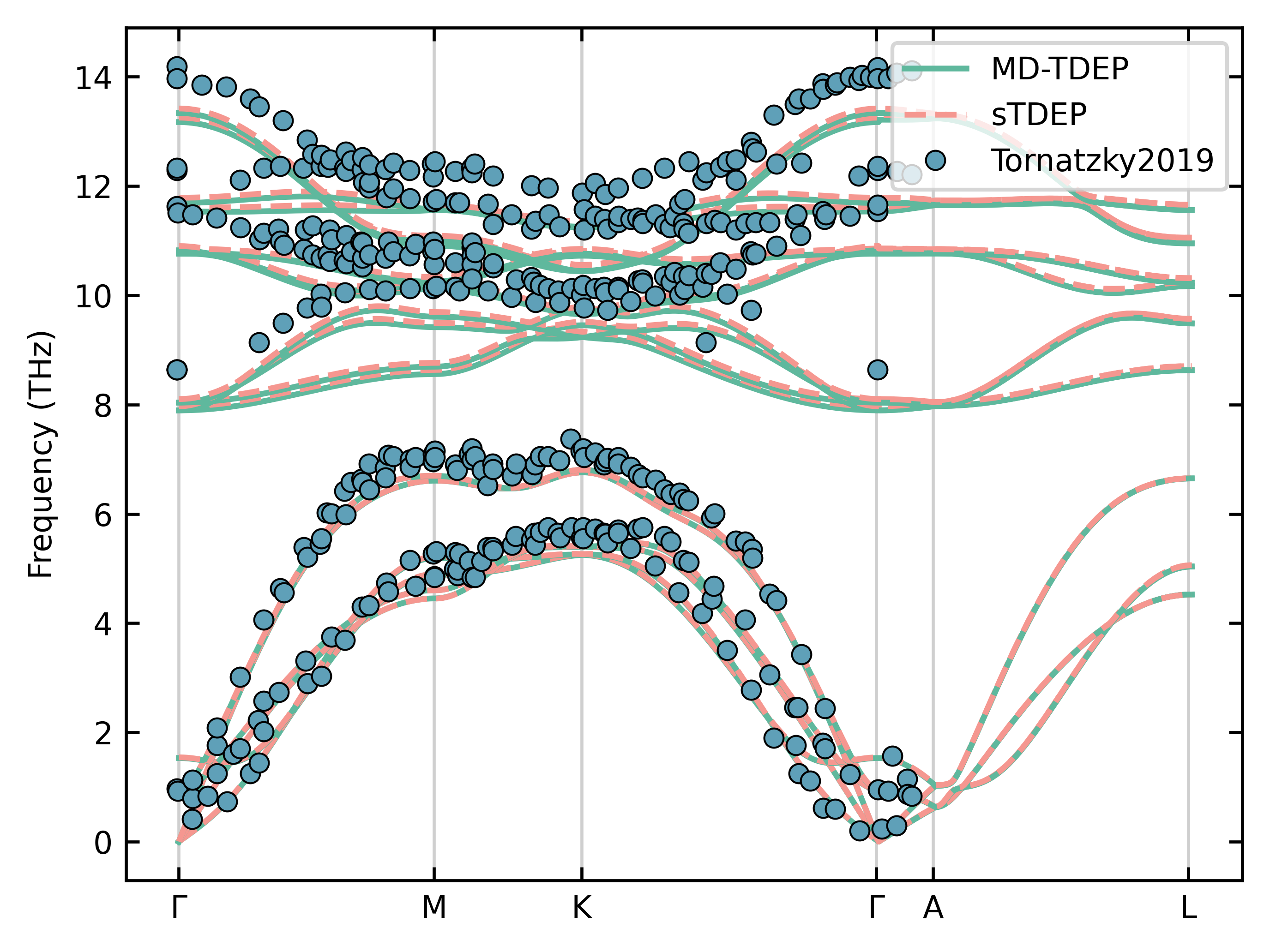}
    \caption{Calculated phonon bands at 300 K with stochastic and MD sampling, and experimental data from Ref.~\cite{Tornatzky2019}.}
    \label{fig:phonons_at_300K}       
\end{figure}

\begin{figure*}[tbh]
    \centering
    \includegraphics[width=0.8\textwidth]{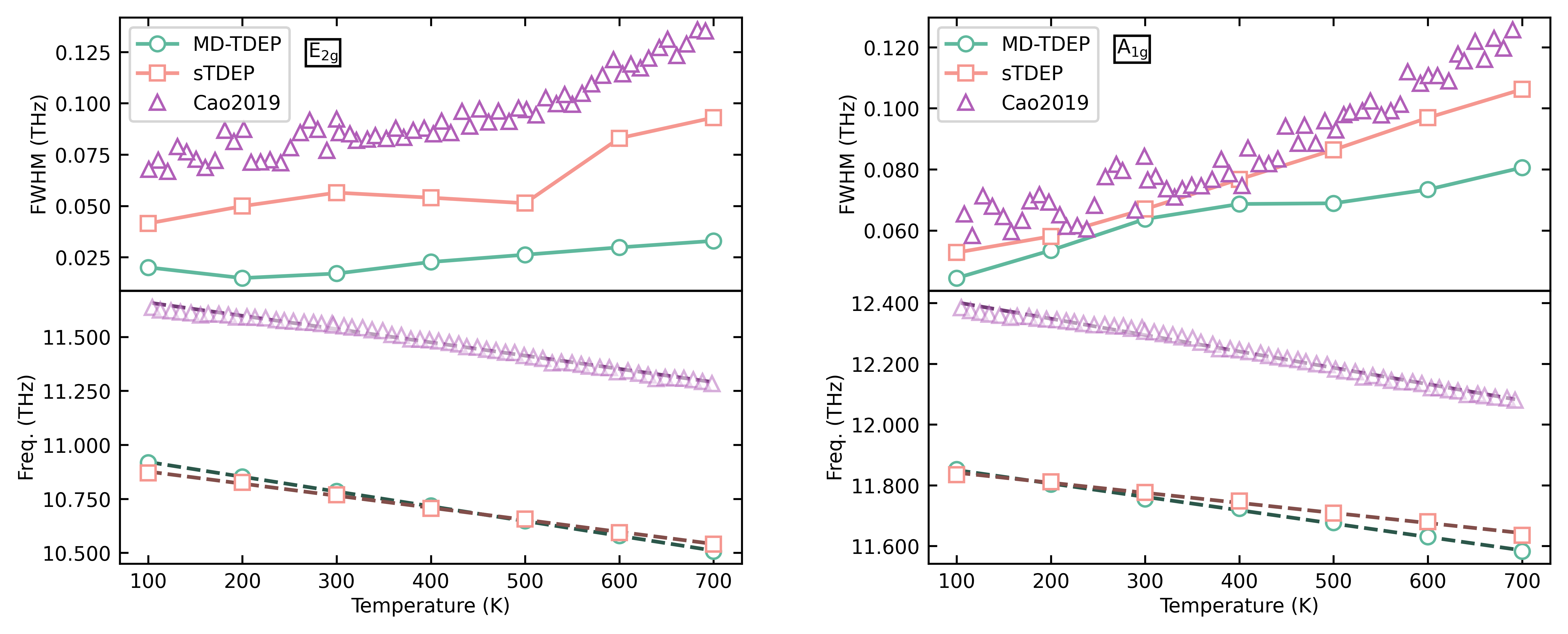}
    \caption{Position and FWHM of the peaks in MoS$_2$ spectral function vs temperature from theoretical calculations (with MD and stochastic sampling) and experimental Raman spectra from Cao and Chen \cite{Cao2019} vs temperature.}
    \label{fig:FWHM_FREQS_vs_T_with_Cao}       
\end{figure*}

\subsection{\label{subsec:vib_props}Vibrational properties}
We use TDEP to extract the IFC2s and IFC3s. The phonon dispersion bands vs. T are reported in \reffig{fig:phonons_vs_T}. As expected, thermal expansion and anharmonicity lead to a general softening of the modes.

In \reffig{fig:phonons_at_300K} we show our phonon bands obtained at 300 K with the MD and stochastic sampling, as compared to the experimental values from Ref.~\cite{Tornatzky2019}. We note that the two sampling methods give very similar phonon bands. Our frequencies are smaller than the experimental ones by a factor which increases with the frequency itself. We attribute this to the DFT approximations used to train the MLIP, as GGA functionals systematically underestimate the phonon frequencies. This discrepancy is also consistent with the slight overestimation of the lattice constants that we present in the previous section.

In agreement with experiments \cite{Wieting1971, Zhang2015, Sahoo2013, Thripuranthaka2014, Lanzillo2013, Livneh2010, Yang2016, Yan2014} and symmetry, the two degenerate \etg{} in-plane (x-y) modes and the \aog{} out-of-plane (z) mode are found to be Raman active in 2H-MoS$_2$. Two other \etg{} degenerate modes are present at low frequency, but they are associated to shear displacement of the layers. Due to their low effect on the electronic polarizability, they show a Raman response orders of magnitude smaller than the first two, and often noisy. Therefore, in this study we focus on the high-frequency modes.

\reffig{fig:FWHM_FREQS_vs_T_with_Cao} presents the FWHMs and centers of the peaks of the spectral function for the two modes, as a function of temperature.

We start by noting that for both modes a temperature increase leads to a red shift in the frequencies. This is the expected behavior as the unit cell expands as a result of increasing temperature. We fit the temperature-evolution with a linear function (intercepts, coefficients and experimental references are reported in \reftable{tab:thermal_coefficients}). Looking at the trends of the frequencies in the full range of temperature, we note that the MD sampling is more sensitive to temperature. Since the same unit cell is used for both the MD and stochastic sampling (the first-order optimizations performed in the second method lead to negligible changes), the difference in the rate of change of the frequencies can simply be ascribed to the quantum phonon occupation in the stochastic method, which starts larger and increases less rapidly in temperature than in the classical case. The rate of change in the occupation is related to that of the mean square displacement which determines how much more anharmonicity is captured upon a temperature increase.  

The \aog{} (out-of-plane) experimental frequency decreases faster vs. T than the theoretical ones. This is consistent with our slight underestimation of the the thermal expansion coefficient for the $c$ axis.

\begin{table}[htpb]
\centering
\begin{threeparttable}
\begin{tabular}{|l|c|c|c|c|}
\hline
\multirow{2}{*}{} & \multicolumn{2}{c|}{\etg} & \multicolumn{2}{c|}{\aog} \\ \cline{2-5}
                   & $\chi_0$ & $\chi_1$ & $\chi_0$ & $\chi_1$ \\ \hline
           MD-TDEP & 1.099 & -6.835  & 1.189 & -4.401 \\ \hline
           sTDEP   & 1.093 & -5.581  & 1.187 & -3.313 \\ \hline
        Cao2019\cite{Cao2019}  & 1.172 & -6.144 &  1.246& -5.389 \\ \hline

\end{tabular}
\end{threeparttable}
\caption{Linear parametrization of the temperature dependent Raman shifts from this study and experimental data in the literature. The $\chi_0$ and $\chi_1$ coefficients are given in $\times$10 THz and $\times$10$^{-4}$ THz/K, respectively.}
\label{tab:thermal_coefficients}
\end{table}

Our widths tend to be smaller than those observed experimentally. This can be ascribed to interactions beyond the third order, and to disorder, inhomogeneous broadening, and measurement noise in the experimental FWHMs. The latter factors complicate a strict quantitative comparison with the purely anharmonic linewidths calculated here. However, our widths follow a similar trend vs. T as in the experiments.
We also note that the stochastic sampling systematically leads to larger FWHMs than the MD sampling. 

\begin{figure}[htb]
    \centering
    \includegraphics[width=\columnwidth]{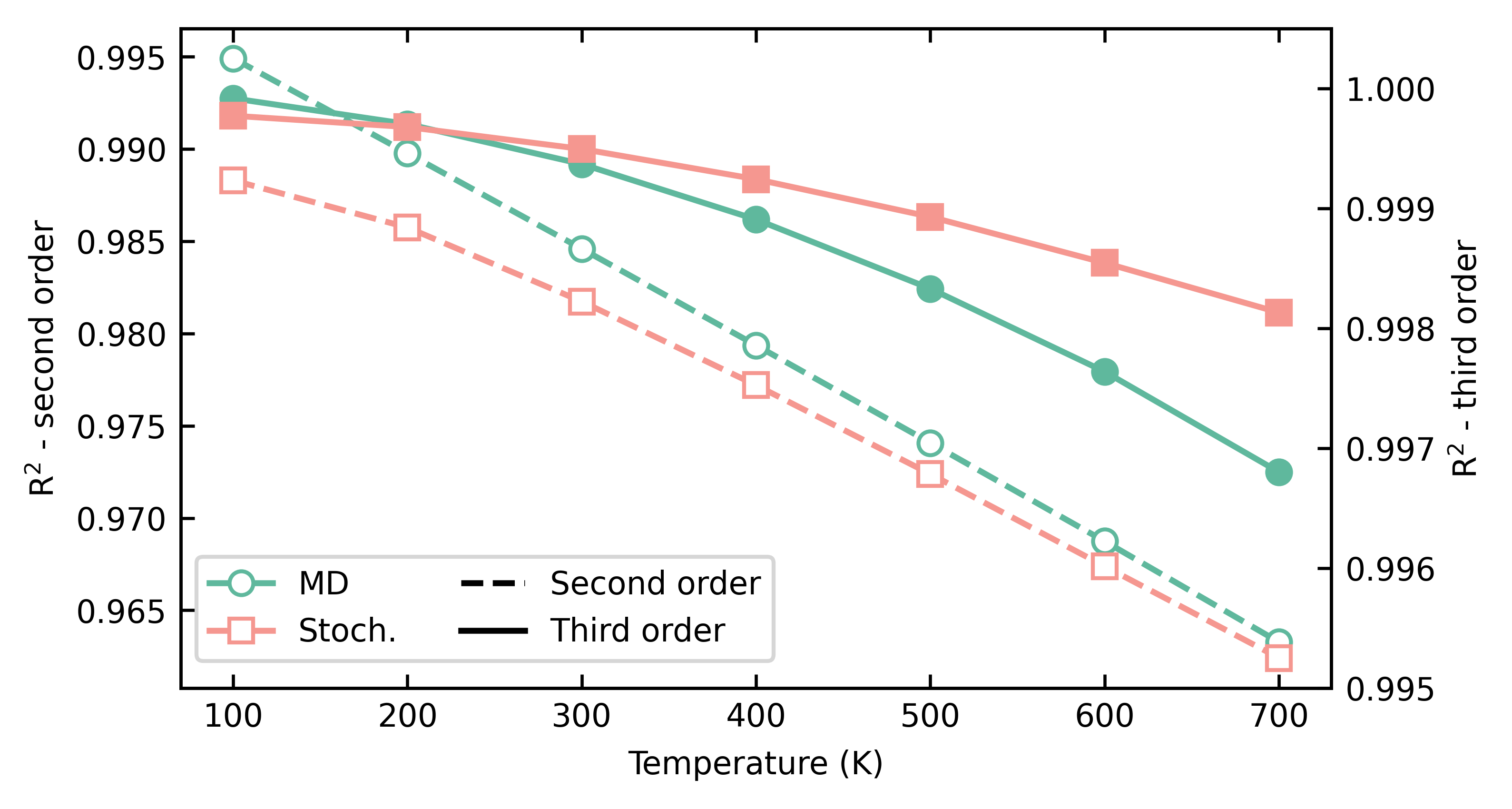}
    \caption{R$^2$ of the TDEP fit of IFCs at the second and third order vs temperature with the MD and stochastic sampling.}
    \label{fig:R2}       
\end{figure}

To better understand the different results between the two methods, we report the fitting R$^2$ at the second and third order in \reffig{fig:R2}. The methods will differ in the amplitude of displacements, and hence anharmonicity order, which they sample: higher anharmonicity pushes down the R$^2$.

At low T the finite quantum occupation ($1/2 + n(T)$) imposes a larger displacement and more anharmonic forces in sTDEP than in the classical MD-TDEP case, leading to lower R$^2$ at all orders for the sTDEP.
The quantum effect persists until roughly the Debye temperature, after which the harmonic fits are very similar.
At high temperature, the MD sampling explores asymmetric displacements,  This is visible for the third-order: the MD R$^2$ decreases faster than the sTDEP one as T increases.

A further analysis of the error, e.g. by including the fourth-order terms, would be beneficial to understand the evolution of the anharmonic properties versus temperature. While with TDEP it is possible to calculate the imaginary part of the self energy at the fourth order evaluated at the harmonic phonon frequency, $\Gamma^{(4)}(\omega_\lambda)$, the calculation of the full fourth-order frequency-dependent self-energy is very involved and not implemented yet. Furthermore, the real counterpart $\Delta^{(4)}(\omega_\lambda)$ is not calculated either.
Therefore, as a first inspection, we calculated the fourth-order contribution to the self-energy, to include it as a rigid translation of the imaginary part:
\begin{equation}\label{eq:approx_self_energy}
\tilde\Sigma_\lambda(\omega)=\Delta^{(3)}_\lambda(\omega) + i\left(\Gamma^{(3)}_\lambda(\omega)+\Gamma^{(4)}_\lambda(\omega_\lambda)\right)
\end{equation}

For the two modes of interest at Gamma, we find that the fourth-order contribution is of negligible amplitude. We believe fourth-order processes may still play a significant role in the shape of spectral function if the full $\Gamma_\lambda^{(4)}(\omega)$ is used and the approximation of rigid translation of the imaginary self-energy is relaxed. This will be re-examined in the future, when the calculation of the full omega-dependent $\Gamma^{(4)}(\omega)$ will be available in TDEP.

The influence of inhomogeneous broadening in the experimental data is compatible with the discrepancy between the ratio of our FWHMs and those from experiments. Indeed, if the same inhomogeneous broadening is applied to both peaks, it has a stronger relative effect on the narrow \etg{} peak, leading to a ratio closer to the experimentally observed one.
Further factors missing from our description can contribute to the broadening of the phonon spectral function, such as their interaction with electrons \cite{Girotto2023} and changes in the many-body electronic screening \cite{Perfetto2023}.
Finally, we mention that our theoretical FWHMs are very sensitive to the numerical and computational parameters, and some results showed very slow convergence (notably the sTDEP FWHMs vs. rc3).

To compare the Raman intensities, we combine eqs. \refeq{eq:raman_scattering_eff} and \refeq{eq:iso_average_1}, to obtain the area of the Raman peak, which we relate to the average Raman intensity by:
\begin{eqnarray}\label{eq:exp_spectrum_sum}
    \int_{\mathrm{peak}}S(\omega)d\omega=
    kN_{\mathrm{peak}}\frac{I_{\mathrm{avg}}^\lambda\left[n_\lambda(T)+1\right]}{\omega_\lambda}
\end{eqnarray}
(here $k$ contains all the other constants, including any experimental factor) assuming that only $N_{\mathrm{peak}}$ degenerate modes contribute to the peak with the same spectral function and Raman response. We extract the intensity as:
\begin{eqnarray}
    I_{\mathrm{avg}}^\lambda=\frac{\omega_\lambda\int_{\mathrm{peak}}S(\omega)d\omega}
    {kN_{\mathrm{peak}}\left[n_\lambda(T)+1\right]}
\end{eqnarray}
Instead of the harmonic frequency $\omega_\lambda$ in the denominator of \refeq{eq:exp_spectrum_sum} and for the Bose-Einstein factor, we use the center of the Raman peak, $\tilde\omega_\lambda$, which corresponds, in our theoretical framework, to the renormalized frequency due to the full-frequency dependent self-energy. Therefore,  we implicitly assume the correction to be small (which is true for the theoretical results). Also, note that while we cannot compute the intensity without knowing $k$, we can compute the ratio between the intensities of different peaks.

We extract the Raman intensity from the experimental spectra of few-layer MoS$_2$ from the work of Sahoo et al. \cite{Sahoo2013}, according to our discussion in \refsect{subsec:Raman_theory}. Specifically, we fit each peak (after removing the baseline) with Lorentzians of the shape:
\begin{eqnarray}
    f(\omega)=\frac{A}{\pi}\frac{\Gamma}{(\omega-\tilde\omega)^2+\Gamma^2}
\end{eqnarray}
with $A$, $2\Gamma$, and $\tilde\omega$ being the area, the FWHM, and the center of the peak, respectively.

\begin{figure}[htb]
    \centering
    \includegraphics[width=0.9\columnwidth]{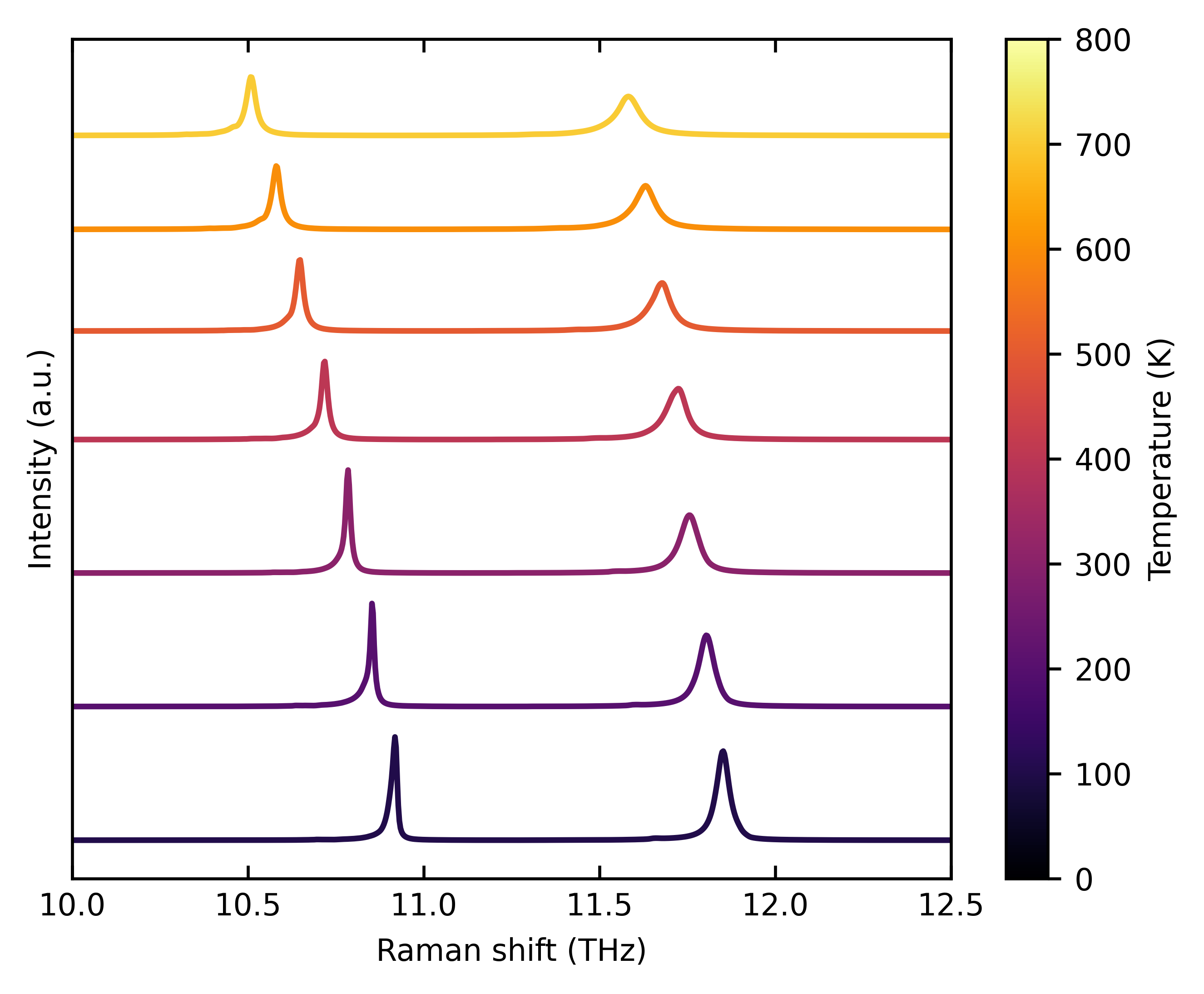}
    \caption{Theoretical isotropically averaged Raman spectra of MoS$_2$ from 100 to 700 K.}
    \label{fig:raman_spectra}       
\end{figure}

\begin{figure}[htb]
    \centering
    \includegraphics[width=0.9\columnwidth]{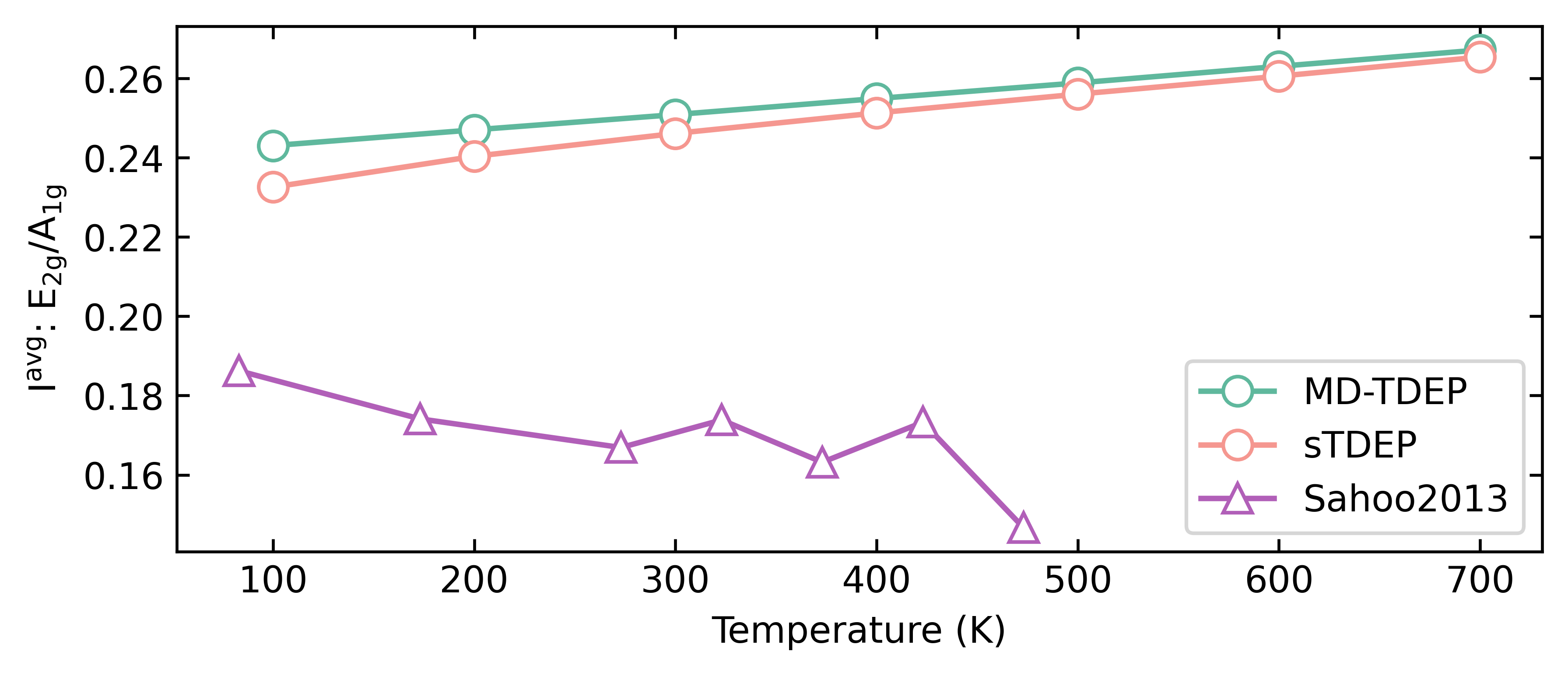}
    \caption{Raman intensity ratio between modes \etg{} and \aog{} in MoS$_2$.}
    \label{fig:raman_intensities_ratio}       
\end{figure}

In \reffig{fig:raman_intensities_ratio} we plot the ratio between the averaged Raman intensity of the \etg{} and \aog{} modes. The ratios between our Raman intensities are of the same order of magnitude as the experimental ones, which, unlike ours, exhibit an overall decreasing non-monotonic trend. Our calculated spectra (\reffig{fig:raman_spectra}) show higher \etg{} peaks than the \aog{}. While this aligns with some experimental reports \cite{Fan2014, Najmaei2012, Najmaei2013}, it contradicts others \cite{Golasa2014, Livneh2015}. This discrepancy likely stems from the fact that relative intensities are highly sensitive to the exciting frequency. In fact, any frequency above the 1.23 eV band gap \cite{Kam1982}, can easily trigger resonant Raman phenomena, while our calculations are performed in the static limit. This highlights the difficulty in comparing different experiments with theory or amongst themselves. For intensities and peak ratios the conditions, substrate, dielectric environment, and laser excitation frequency and polarization must be known very precisely.

\section{\label{sec:conclusions}Conclusions}
We calculate the Raman spectra of 2H-MoS$_2$ from 100 to 700 K with the TDEP approach, so as to include finite-T effects on the position and width of the Raman peaks. We generate a representative dataset and use it to train an accurate MTP model, which accelerates the determination of the thermalized cell parameters, and the thermodynamic sampling. 
We compare two different sampling methods: classical MD simulation of the atomic trajectories through phase space, and stochastic generation of structures from an effective canonical ensemble according to thermally renormalized phonons. 
We obtain cell parameters in very good agreement with experimental measurements. Our phonon band dispersions at room temperature also agree with those from experiments, confirming that our method gives a good description of the lattice dynamics at the harmonic level, including the renormalization coming from the interaction between phonons. We calculate the phonon self-energy with third-order contributions, obtaining phonon lifetimes which compare well to the experimentally observed ones, in absolute terms and for their temperature dependence. These results could be improved by accounting for contributions of phonon processes at higher-order in a complete way (full frequency-dependent phonon self-energy at the fourth order), as well as electron-phonon interactions and electronic screening modulations due to many-body effects. The Raman intensities obtained from the DFPT dielectric calculations are compared with those extracted from the experimental spectra. They are of similar order of magnitude, but depend very strongly on the experimental (and numerical) conditions. We ascribe at least part of this discrepancy to most measurements being performed in the resonant Raman regime. Future experimental works with exciting light below the band gap might confirm (or contradict) our resonance-free results.   

\section{Acknowledgements}
The authors thank F. Knoop, O. Hellman and J.P. Alvarinhas for insightful discussions. S.L. acknowledges funding from Horizon Europe MSCA Doctoral network grant n.101073486, EUSpecLab, funded by the European Union.
M.J.V. acknowledges funding by the Dutch Gravitation program
“Materials for the Quantum Age” (QuMat, reg number 024.005.006), financed by the Dutch Ministry of Education, Culture and Science (OCW).
Simulation time was provided by 
the Consortium d'Equipements de Calcul Intensif (FRS-FNRS Belgium Grant No. 2.5020.11).
EuroHPC-JU award EHPC-EXT-2023E02-050 on MareNostrum 5 at Barcelona Supercomputing Center (BSC), Spain;
and by the Lucia Tier-1 of the F\'ed\'eration Wallonie-Bruxelles (Walloon Region grant agreement No. 1117545).

\bibliographystyle{unsrt}
\bibliography{references_fixed}
\end{document}